\documentclass[10pt,letterpaper]{article}
\usepackage[top=0.85in,left=2.75in,footskip=0.75in]{geometry}

\usepackage{amsmath,amssymb}

\usepackage{changepage}

\usepackage[utf8x]{inputenc}

\usepackage{textcomp,marvosym}

\usepackage{cite}

\usepackage{hyperref}

\usepackage[right]{lineno}

\usepackage{microtype}
\DisableLigatures[f]{encoding = *, family = * }

\usepackage[table]{xcolor}

\usepackage{array}

\newcolumntype{+}{!{\vrule width 2pt}}

\newlength\savedwidth

\raggedright
\usepackage[aboveskip=1pt,labelfont=bf,labelsep=period,justification=raggedright,singlelinecheck=off]{caption}
\renewcommand{\figurename}{Fig}

\makeatletter
\renewcommand{\@biblabel}[1]{\quad#1.}
\makeatother

\date{}

\usepackage{lastpage,fancyhdr}
\usepackage{graphicx}
\fancyheadoffset[L]{2.25in}
\fancyfootoffset[L]{2.25in}

\newcommand{\rank}[1]{r(#1)}
\newcommand{\correggi}[1]{}
\newcommand{\rca}{R} 
\newcommand{\trca}{RCA} 
\newcommand{\exm}{E}
\newcommand{\lpd}{logPRODY}
\newcommand{\cpx}{Q}
\newcommand{\gdp}{Y}
\newcommand{\nrca}{\tilde{R}} 
\newcommand{\theplane}{RCLP}
\newcommand{\asymptoticzone}{asymptotic zone}
\newcommand{\asymptoticmarket}{asymptotic market}
\newcommand{\asymptotic}{asymptotic} 

\begin{document}
\vspace*{0.2in}

\begin{flushleft}
{\Large
\textbf\newline{The Complex Dynamics of Products and Its Asymptotic Properties} 
}
\newline
\\
Orazio Angelini\textsuperscript{1*},
Matthieu Cristelli\textsuperscript{1},
Andrea Zaccaria\textsuperscript{1},
Luciano Pietronero\textsuperscript{1,2},
\\
\bigskip
\textbf{1} ISC-CNR, Institute for Complex Systems, Rome, Italy
\\
\textbf{2} Physics Department, Sapienza University of Rome, Rome, Italy
\\
\bigskip

%
%





* angelini.orazio@gmail.com

\end{flushleft}
\section*{Abstract}
We analyse global export data within the Economic Complexity framework. We couple the new economic dimension Complexity, which captures how sophisticated products are, with an index called logPRODY, a measure of the income of the respective exporters. Products' aggregate motion is treated as a 2-dimensional dynamical system in the Complexity-logPRODY plane. We find that this motion can be explained by a quantitative model involving the competition on the markets, that can be mapped as a scalar field on the Complexity-logPRODY plane and acts in a way akin to a potential. This explains the movement of products towards areas of the plane in which the competition is higher. We analyse market composition in more detail, finding that for most products it tends, over time, to a characteristic configuration, which depends on the Complexity of the products. This market configuration, which we called asymptotic, is characterized by higher levels of competition.


\section*{Introduction}
In this paper we report the results of an analysis that has been triggered by recent developments in the field of Economic Complexity. The proposal of the Fitness and Complexity measures \cite{Tacchella2012}, which capture respectively the advancement of the productive system of a country and the level of sophistication of a given kind of product found in the export market, allows for the extraction of useful information from countries' export data. This enabled a novel approach that compared monetary metrics, such as Gross Domestic Product per capita (GDPpc or $\gdp$ from now on), with non-monetary metrics, such as Fitness and Complexity\cite{Cristelli2013}. The interplay between these quantities reveals a wealth of information about economic phenomena. With this method, we explore the dynamics of the Complexity ($\cpx$ from now on), comparing it to logPRODY, a monetary metric defined as the weighted average of the GDPpc's of a product's exporters\cite{Hausmann2007}. This lets us gather insight into the dynamics of the global export markets for different products, and find a number of relevant regularities in their behavior.\newline

The Complexity measure stems from a promising body of research. The hypothesis underlying the definition of this non-monetary metrics is that counties' competitiveness is connected with their \emph{capabilities}\cite{Hausmann2009,Dosi1988,Lall1992,Teece1994}, which are non-exportable and hard-to-measure features that allow the production of sophisticated goods\cite{Hidalgo2007}. For example, property rights, regulation, educational system, infrastructure, the specific know-how of an industry, etc., could all be considered to be capabilities. In principle, though, there can be any number of capabilities with any relative importance, and it is impossible to ``a priori'', or normatively, determine them. This makes it practically impossible to measure them directly. A recent body of research focused on leveraging the economic output of a country as a proxy for its capabilities, by looking at which products it is capable of exporting \cite{Hidalgo2007,Hausmann2009,Tacchella2012,Cristelli2013a,Zaccaria2014}.
The fact that a country is capable of exporting a given product $p$ signals that its production system is competitive enough to stand out in the global market for $p$. This, in turn, implies that it has the necessary capabilities to produce $p$. In other words, products encode information on the (up to now) elusive capabilities. How to infer the presence of capabilities from export data? World trade flow data is representable as a bipartite network, where country $c$ exporting product $p$ is represented as a link between $c$ and $p$. This allows a bottom-up approach enabling us to exploit the network structure in order to measure properties related to the capabilities from data regarding economic output. Conceptually this body of literature is doing something similar to PageRank\cite{Page1998,Caldarelli2013}, since it leverages topological properties of the network in order to measure properties of the nodes. The first attempt is due to Hidalgo and Hausmann\cite{Hausmann2009}, followed by the aforementioned Fitness and Complexity measures. The adjacency matrix of the export bipartite network is nested\cite{Tacchella2012}. This means that there are some countries that export almost all the products, and some products that are exported only by the few most diversified countries, that also happen to usually be the richest ones from a monetary point of view. On the other hand, the few products exported by poorly diversified countries are also exported by almost everyone else. A finding in itself, that drives the subsequent analysis, is that the evolution of an economy in time is characterized by increasing diversification, rather than specialization\cite{Cadot2011}. The situation is analogous to what is found in some biological systems\cite{Bascompte2003}, namely that the network is \emph{nested}, and this offers a natural way to sort products in terms of increasing \emph{Complexity} and countries in terms of \emph{Fitness}\cite{Cristelli2013a} (indeed a similar algorithm has actually been applied to biological networks\cite{Dominguez-Garcia2015}). The definition and calculation techniques for Fitness and Complexity are discussed at length in the Methods section.\newline

Comparing non-monetary metrics to monetary ones promises to be a particularly insightful approach. As an example, a fruitful way to look at the Fitness non-monetary metric is to compare it with the monetary Gross Domestic Product per capita. By representing countries with their position on the Fitness-GDPpc plane, it has been possible to apply techniques of dynamical systems in order to model their coupled motion\cite{Cristelli2013}. In some cases, the motion is sufficiently regular to enable predictions on the future change in Fitness and GDPpc.\newline

In light of this promising approach in literature, we set out to analyze Complexity, first by identifying a suitable monetary metric to compare it with, and then looking at its dynamics with a method similar to the one used for Fitness. We considered various possibilities for the monetary index, the most relevant being product prices, value added, and two metrics called PRODY\cite{Hausmann2007} and Sophistication\cite{Lall2006}.\newline 

Prices are very impractical to use, since it is difficult to collect reliable data about them, and they fluctuate in response to a large number of variables that might have little to do with their Complexity. We tried extrapolating prices from the BACI dataset (see Methods for information on the datasets), which contains, for each product and couple of countries, the unit quantity exported and the price paid for it in dollars. Unfortunately the data is partly unreliable because of non-standardized customs' procedures, and also because a given class of products allows very different pricing in it. Additionally, given how heterogeneous products are, there is no universal definition for the quantity of goods being exported, making it difficult to compare across different kinds of goods. Because of this, the only relevant information that can be extracted for the data is the relative change in time of prices for highly aggregated categories. This results in extremely noisy data, with which we are not able to reliably explore any hypothesis. Value added, being defined as the difference between production cost and selling price, retains all the problem prices have, adding to them the problem of reliably measuring the production cost. Since it is an extensively studied quantity, though, it is still possible to find reliable data, but it is very aggregated. The main datasets are the OECD Trade in Value Added\cite{OECDtvad2016} data and the World Input Output Database\cite{WIOD2016}. Both contain a handful of broad categories, with which we couldn't perform the kind of analysis presented in this paper.

Finally, PRODY and Sophistication appear to be good candidates for our study. The quantities used to calculate them are widely available and reliable. They have already been used to classify products and make predictions\cite{Lall2006,Hausmann2007,Jarreau2012,Ghani2011,Jarreau2009}, and they depend on the GDPpc's of exporting countries, so they look like a natural counterpart to Complexity as GDPpc is to Fitness. We slightly modifed PRODY by defining $\lpd$, which looks to us like a more natural measurement. What $\lpd$ represents is a weighted average GDPpc of the exporters of a certain product (the exact definition of PRODY and logPRODY is reported in the Methods section). The biggest the share of a certain product $p$ in the total export of a country $c$, the more $c$ is weighted in the average that defines $\lpd_p$. logPRODY can be interpreted as a proxy for the composition of the market for a given product, signaling what is the mean income of countries export it. A change in logPRODY is linked to a change in the underlying composition in the set of countries that export a given product and, as we will see, these changes can be modeled in analogy with dynamical systems. This is essential, because high correlation is expected (and observed) between logPRODY and Complexity: high-$\cpx$ products are exported by high-Fitness countries, which are usually the richest. By just using traditional parametric regression techniques, we would not be able to gather much more information than this simple fact. Instead, the technique we used allows us to check what happens, for example, to products that don' follow, or stop following, the trivial correlation trend.\newline

What we actually observe in this analysis is the motion of the products on the ranked Complexity $\rank{\cpx}$ vs. ranked logPRODY $\rank{\lpd}$ plane (called \theplane\ from now on). Note that the ranking procedure is simply an assignment of a 0 score to the product with the lowest yearly value and a 1 score to the highest, while the other products are equally spaced in this interval; see Methods on why the ranking is our choice in analysis. Analysing time displacements allows us to detect non-trivial interactions between the two quantities, that would be hard to detect with other methods. We also assign a value of competition on the export market, as measured by the Herfindahl index (explained further on), to each area of the plane, as it seems competition has some explanatory power for the motion we observed.

\section*{Results}
Here we will illustrate the findings of our analysis. This section is divided in paragraphs illustrating respectively the motion we observe, a model we suggest to explain its cause, the concept of \asymptoticmarket\ that we will introduce further on, and finally a minor discrepancy between the model and the observation, that is explained outside of the model. The analysis is conducted on two different datasets, which we call BACI and Feenstra respectively, and are detailed in the Methods section. We want to stress that in all cases the two datasets give results that are consistent with each other.\newline

\begin{figure}
\centering
    \includegraphics[width=\textwidth]{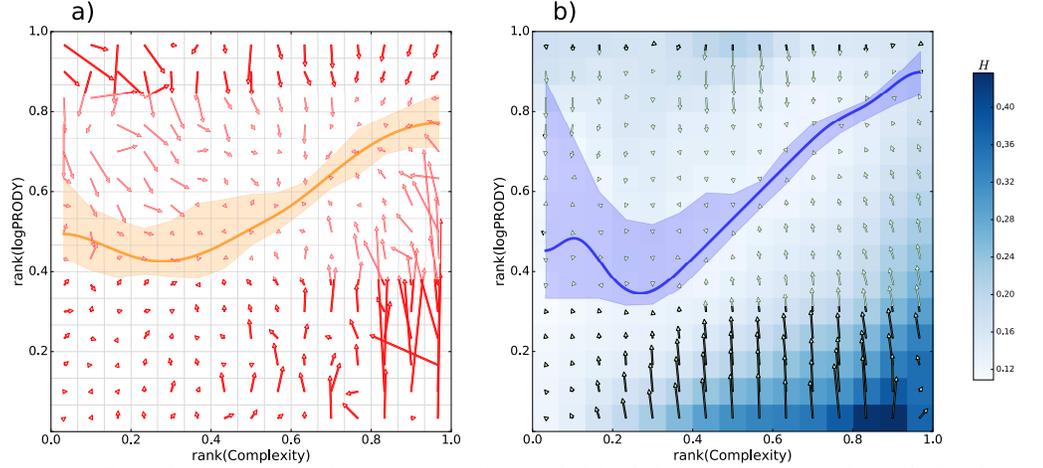}
    \caption{\textbf{Panel a).} In red, the average velocity field of the products on the \theplane, $\vec{v}$. Average velocities always point toward the \asymptoticzone, marked by the orange line, which is calculated by finding the per-column minima of the velocity field after applying a kernel regression on the field. The result of a bootstrap, in orange, indicates the 5\% confidence interval on the values. $\vec{v}$ indicates that points that stray away from the \asymptoticzone\ tend to go back to it.  Low velocities do not mean that products do not move, but rather that they move away from their box in all directions. The motion develops mainly across the vertical direction, with points randomly diffusing away from the \asymptoticzone, and then moving back to it with regularity. \textbf{Panel b).} Average Herfindahl per box, i.e. the $H$ field (in blue color), with its gradient calculated according to Eq. \ref{eq:hfield-grad-def} (green arrows). Higher values of $H$ signal that there is less competition, on average, for products in that box. The blue line is a kernel regression of the minima of the Herfindahl field per column, with the blue colored area showing the 5\% confidence interval on the value, calculated with a bootstrap. Notice its position is compatible with that of the velocity field minima. The $H$ field acts in a way akin to a potential, as the products move towards its minima. This strongly suggests that changes in the vertical position of products is due to changes in the underlying markets. All the vectors in both panels are to 1:1 scale with each other and with the plot's axes.}
    \label{fig:quiver-rank_rank}
\end{figure}

\textbf{Kinematics.} To gather information on the dynamics of the products' Complexity, we can look at the ranked Complexity ($\rank{\cpx}$) vs. ranked logPRODY ($\rank{\lpd}$) plane. We find that, as expected, there is a strong relation between $\rank{\cpx}$ and $\rank{\lpd}$: most of the points tend to gather on the diagonal of the plane (for the dataset provided by Feenstra et al. in \cite{Feenstra2005}, see Methods for details on the data used), with a dramatic change in the density of points. The Spearman correlation coefficient between $\rank{\cpx}$ and $\rank{\lpd}$ is 0.72. From the most populated to the least populated area of the plane, density changes by about 3 orders of magnitude (A plot of the datapoints is shown in the Supplementary Information addendum). Next, we define the vector field $\vec{v}$ representing the average velocities on said plane, see \figurename\ref{fig:quiver-rank_rank}. To calculate $\vec{v}$, we divide the plane into a  grid of boxes. Each arrow in \figurename\ref{fig:quiver-rank_rank}, panel a), represents the average of all 1-year products' displacements whose starting point is found in the corresponding grid of the box; refer to Methods section for further information on this regression technique. The motion resembles a laminar flow: points on the plane tend to diffuse out of the grid boxes with a certain average velocity, which is a function of the position of the box. The average velocities always point toward a line that is slightly skewed from the diagonal, which we will call the \emph{\asymptoticzone}, highlighted in orange in \figurename\ref{fig:quiver-rank_rank}. The velocities are higher the farther away one gets from the \asymptoticzone. Lower velocities on the \asymptoticzone, however, do not mean that products do not move; the low average velocity here means instead that points generally move away from the \asymptoticzone\ in all different directions. In general, there is an ordered movement towards the \asymptoticzone, shown in \figurename \ref{fig:quiver-rank_rank} panel a),  and a random movement away from the \asymptoticzone, which cannot be pictured via the average velocity  field $\vec{v}$, since the random component averages to zero. The shape of $\vec{v}$ holds at different time scales, i.e. if we look at the displacements over time intervals ranging from 1 to 10 years, and if we decrease or increase the number of boxes in the grid up to the point where there is at least a number of order ~5 points per box (see Supplementary Information). The areas away from the \asymptoticzone, though, do not get emptied over time. What seems to happen is that points tend to diffuse randomly away from the \asymptoticzone, but to fall back to it with regularity, their typical return trajectories being captured by $\vec{v}$. We attribute the movement away from the \asymptoticzone\ to contingencies that change the export market for a product and move its position away in a random direction. The sum of the orderly movement towards the \asymptoticzone\ and the random movement away form it returns a picture that is self-consistently stationary on the \theplane, meaning that it keeps its characteristics unchanged over time. Further calculations in support of this stationarity are shown in the Supplementary Information. The velocity field can appear non-stationary because it shows where a product will go, on average, after it is found in a given box, but it does not describe how products enter the boxes. Calculating this "inward" velocity (shown in Supporting Information), a a clear pattern of diffusion out of the equilibrium diagonal emerges. Changes in density of points per box over time (also found in the Supporting Information) are negligible.
This stationary behavior is consistent across all datasets examined. These findings suggest that points in the highly populated \asymptoticzone\ are in a sort of equilibrium situation (on which we will elaborate further on), where the average logPRODY is at the ``required'' value with respect to the Complexity of a product. Points occasionally diffuse away from it, but whenever a point is away from the \asymptoticzone, meaning that logPRODY is too high or too low with respect to Complexity, it tends to move towards it. Products farther away from the \asymptoticzone\ also tend to move considerably faster than those lying close to it. This is the observation that led us to the conclusion that there is an \asymptotic\ value of logPRODY  for a product, which depends on its Complexity.\newline

\textbf{What drives the motion.} What is the process driving the motion on this plane, i.e. what causes the products to diffuse away from the \asymptoticzone, and then go back to it? The answer can be sought by looking at what happens to the export market for a given product in time, and how does this affect the motion and the position of the product on the plane.

We characterize the market by using the Herfindahl index\cite{Rhoades1993,WAKelly1981}:
\begin{equation}
H_p = \sum_c \left( s_{cp} \right)^2 ; \qquad s_{cp} = \frac{\exm_{cp}}{\sum_c \exm_{cp}}
\end{equation}
where $s_{cp}$ is the share of country $c$ in the global export market for product $p$. A sum of the squared market shares of each country, the Herfindahl index equals 1 when the market is a monopoly, and decreases as the competition increases. Calculating the average Herfindahl index per box on the $\rank{\cpx} - \rank{\lpd}$ plane reveals another pattern. Since such a calculation returns a scalar field sampled at discrete intervals, we call the result the $H$ (Herfindahl) field. The scalar field obtained is quite irregular in the high frequencies, therefore we looked at the lower frequencies of the field by smoothing it with a Gaussian kernel convolution. Superimposing the velocity field $\vec{v}$ shows not only that products drift toward areas of lower Herfindahl index (higher competition), but that the minima of the $H$ field lie in the \asymptoticzone, where the velocity field's modulus is minimum as well, as shown in \figurename\ref{fig:quiver-rank_rank}. This suggests that $H$ is acting in a way similar to a potential\cite{Goldstein2007}. The motion of producs on the RCLP is the result of actions of their producers and consumers. A product moving from low-competition to a high-competition is, usually, the result of previously less relevant producers increasing their market share in a low-competition market, thus making it, by definition, more competitive. So what we observe can be interpreted as producers actually making the obvious choice: ceteris paribus, it's easier to enter a "blue ocean"\cite{kim2005blue} than a "red ocean". The interesting finding is that when competition on a product increases, this is associated on the RLCP with movement towards the "asymptotic zone". The opposite - products moving or staying into low-competition areas when their market competition increases - does not seem to happen. In other words, there seems a tendency for low-competition products to have a logPRODY different from what their Complexity would suggest (or vice-versa).\newline

Note that the derivatives of the Herfindahl field are proportional to the average  velocities directly, and not to accelerations. We can explain this result as the outcome of our averaging process, if we interpret the derivatives of $H$ as forces $\vec{F}$ acting on the products in the boxes defined by the grid on the plane. Each product spends an interval $\Delta t$ in a given box, and exits the box at the end of the interval with a velocity $\vec{v}$. This velocity can be decomposed into two components. One, $\vec{v}_{H}$, is proportional to $\vec{F} \Delta t$, and is due to the action of $H$. The other, $\vec{v}_{O}$ depends on the other degrees of freedom of the system. When we average all the outgoing velocities in a box, we obtain:
\begin{equation}
 \langle \vec{v} \rangle = \frac{1}{N} \sum_i^N \left( \vec{v}_{i, H} + \vec{v}_{i, O} \right) = \langle \vec{v}_{i, H} \rangle + \langle \vec{v}_{i, O} \rangle.
\end{equation}
where $\langle \vec{v}_{i, H} \rangle \propto \vec{F} \Delta t$. If $\langle \vec{v}_{i, O} \rangle$ is reasonably small, this line of reasoning explains our finding that $\langle \vec{v} \rangle$ is correlated to $\vec{F}$.\newline

We calculate the gradient of $H$, see \figurename\ref{fig:quiver-rank_rank} panel b, and find that the following relation holds well:

\begin{equation}
  \vec{v} \simeq - k_{x} \frac{\partial H}{\partial x} \vec{x} - k_y \frac{\partial H}{\partial y} \vec{y} \equiv -\vec{\nabla}_{k} H \label{eq:hfield-grad-def}
\end{equation}

where, for readability reasons, we refer to the Complexity direction on the plane as $\vec{x}$ and the logPRODY direction as $\vec{y}$. To check the validity of equations we evaluate a linear regression between the vertical and horizontal components of the fields. The $k_x, k_y$ coefficients in this formula are the first-order coefficients of this regression; zero-order coefficients are set to zero by hypothesis. The gradient of $H$ has to be scaled differently along the two components to correctly reproduce the velocity field $\vec{v}$, an operation we defined with the symbol $\vec{\nabla}_{k}$.
The concordance between Eq. \ref{eq:hfield-grad-def} and the data is further discussed in the Supplementary Information addendum. The lower value for the $x$ axis is caused mainly by big outliers in the lower right corner of the plane, where $\vec{v}$ has very high velocities compared to $H$'s gradient. Another feature is the difference between $\vec{v}$ and $-\vec{\nabla}_{k} H$ in the upper left quadrant of the plane: here Complexity tends to increase slightly, but the model does not predict this phenomenon, and we will discuss the issue further on. Eq.\ref{eq:hfield-grad-def} holds at different time scales, from 1 to 10 years time interval, and at different resolutions of the grid (See Supplementary Information). It suggests very strongly that the process underlying the motion on the plane is driven by a change in the market for the represented products. The number of exporters of a given product, when it is in an area of the plane with low average $H$, tends to regularly increase, and this seems to be the driving force changing $\lpd$ and $\cpx$. Another clue is that velocity is the lowest where $H$ is at the minimum for each column. \\
There is a further argument towards the hypothesis that motion on the plane is driven by a change in the set of exporters of a product. The motion of products on the plane from one year to another is too fast for it to be caused just by inter-year change in the GDPpc and Fitness values of exporters. These quantities change too slowly (a few percent at most, yearly) with respect to the speed of products, that can jump most of the ranking in a single time interval. A change in the set of exporters of a product is thus necessary to cause the observed trajectories. \newline

The finding of Eq. \ref{eq:hfield-grad-def} suggests a scheme where most of the products are at their \asymptoticmarket, with the logPRODY reflecting their Complexity. Their moving away from the \asymptoticmarket\ is unpredictable and accompanied, on average, by a decrease of the competition on the market for a given product. Their return to \asymptoticmarket\ is instead much more regular and evident (see \figurename \ref{fig:quiver-rank_rank} panel a) and corresponds to an increase in competition. This mechanism seems to be reliable enough that we can describe the motion on the plane with an equation that connects velocities to competition, and we can confidently say that vertical motion on the plane corresponds to shifts in the market composition.\newline

\begin{figure}
\centering
    \includegraphics[width=\textwidth]{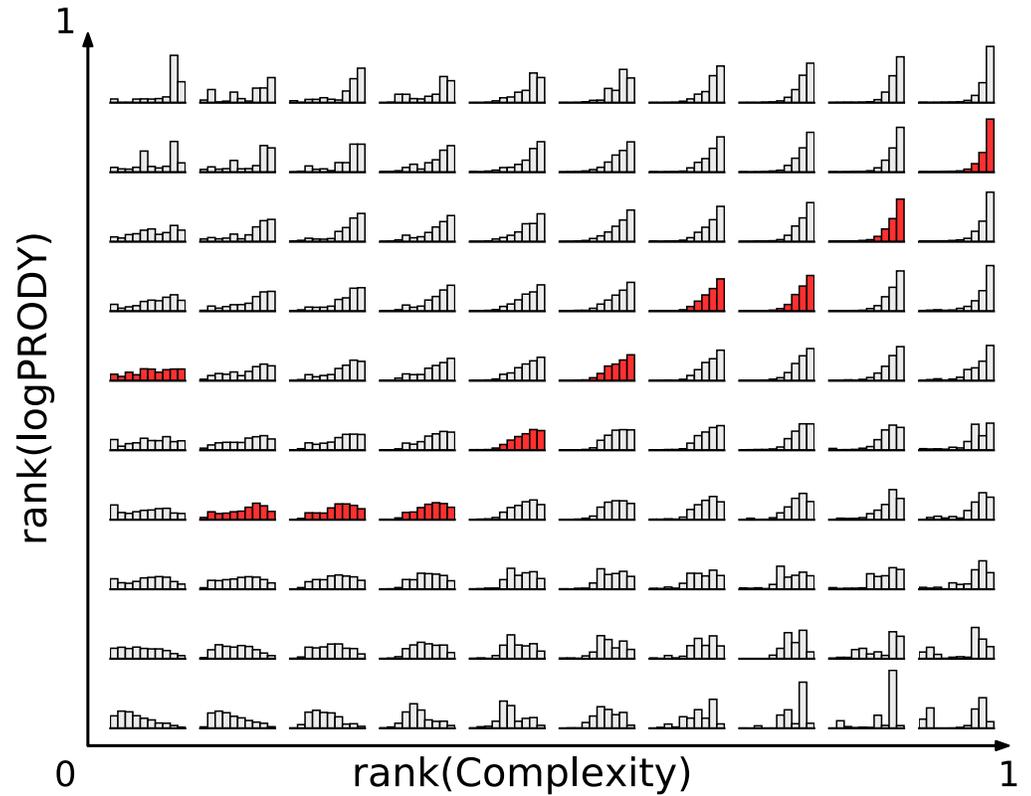}
    \caption{\textbf{Average \trca\ weights per box}. For each box $b$ we plot a histogram showing the average $\nrca$ values of countries exporting the products contained in $b$. Each bar of the histograms shows the average \trca\ of countries with a Fitness value between two consecutive deciles of the Fitness distribution. We remind that $\nrca$ represents the share of a product in a country's total exports. The distributions on the minima of $H$, which correspond to the \asymptoticmarket\ shape of the market, go from flat for the lowest Complexity level to markedly peaked on high Fitness for the highest Complexity, and are highlighted in red. Most of the high-Complexity products are found between the \asymptoticzone\ and the top of the plane, and they show a distribution of comparative advantage that is markedly peaked on high-Fitness countries. More interesting is the case of low-Complexity products, that are mostly found between the \asymptoticzone\ and the bottom of the plane. Here the \asymptoticmarket\ market shape is either rather flat or gently peaked on low-Fitness countries.}
    \label{fig:marketvectors}
\end{figure}

\textbf{Definition of \asymptoticmarket.} We showed that the motion on the plane of products is related to the state of their markets. But what are the actual changes in the markets? To answer this question, we consider our interpretation of logPRODY as a synthetic indicator of the market's composition, as it represents a weighted average of GDPpc's. By analyzing the $\nrca$
weights, as defined in Eq. \ref{eq:logprodydef}, we can have a more detailed description of what the market looks like. We remind that the $\nrca_{cp}$'s are the weighs in the logPRODY. They amount to the \trca\ normalized to one, and they are therefore proportional to the ratio of $p$'s export of country $c$ to all of $c$'s export. The definition is: $\nrca_{cp} = \frac{\rca_{cp}}{\sum_c \rca_{cp}}$. In order analyse the $\nrca$ weights, we produced \figurename \ref{fig:marketvectors}. For each box we calculated a histogram that shows the average $\nrca$ values of countries exporting the products therein contained. Each bar of the histograms shows the average \trca\ of countries with a Fitness value between two consecutive deciles of the Fitness distribution. So the first bar represents the average \trca\ of countries which are exporting the products in the box and have Fitness lower than the first decile; the second bar shows the \trca\ of countries with Fitness between the first and the second decile, and so on. $\nrca_{cp}$ is proportional to the comparative advantage country $c$ has in making product $p$. In short, we are representing the state of the market for products in a certain box in terms of the comparative advantage countries have in making them.\\
The result is shown in \ref{fig:marketvectors}, with the red highlighting the minima of the $H$ field for each column. One can clearly see that the distributions on the minima of $H$ go from flat for the lowest Complexity level to markedly peaked on high Fitness for the highest Complexity. These distributions tell us something about the shape of the market for products found in a given box: the $\nrca$ values, as discussed above, tell us the relative importance of a certain export for the corresponding country. While the high complexity products show a peak of \trca\ on the countries with the highest Fitness, the lowest complexity products exhibit more variation. Most of the low complexity products are found between the lower edge of the plot and the line drawn by the minima of the $H$ field. In this area the market shape shows either a gentle peak around the lower end of the Fitness spectrum or a flat line. This denotes two different market regimes for low-complexity products. One is held by the majority of products, where the comparative advantage is slightly higher for low-Fitness or mid-Fitness countries (but without the sharp peak seen in high-Complexity products), and one at the minimum of $H$, in which all countries tend to develop about the same comparative advantage, causing the minimum in $H$. The velocity in this area is very low and points generally from the mildly peaked distributions towards flatter ones. \newline

All these observations point to the fact that there seems to be a characteristic configuration of the export market for a product, that depends on its Complexity levels. We call this the ``\asymptoticmarket\ ". High-Complexity products show a sharp peak of comparative advantage on high-Fitness countries, and very low comparative advantage on low-fitness countries. This is expected, as Complexity is bounded by the lowest Fitness found among exporting countries. Low-Complexity products, on the other hand, seem to have two possible configurations: either a mild peak, or a flat distribution. There is a mild, but consistent, tendency to go from slightly peaked towards flat. High competition is associated, on the \theplane\ , with having a market in the asymptotic state, and deviations from it are associated with lower levels of competition. \newline

\begin{figure}
\centering
    \includegraphics[width=.5\textheight]{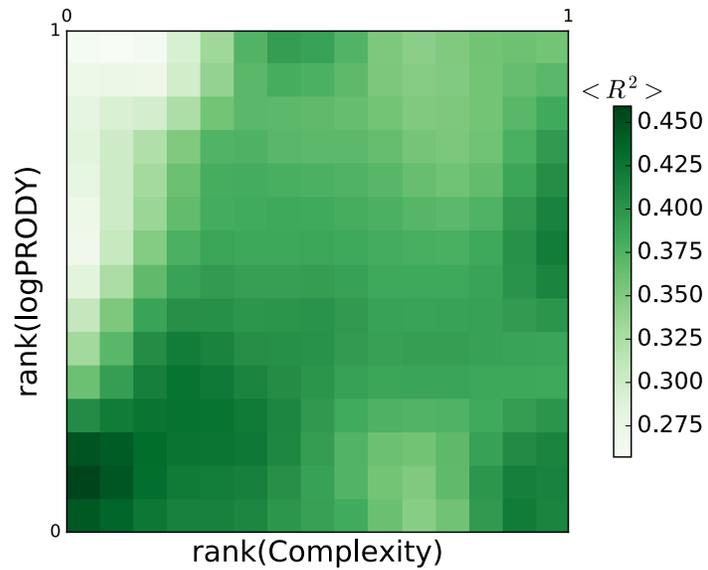}
    \caption{\textbf{Beyond Herfindahl: Fitness and GDP correlation.} We try to explain why the model does not predict positive velocities along the Complexity axis in the top left part of the plane (see \figurename \ref{fig:quiver-rank_rank}). We look at the exporter set of products. For each product $p$, we take its top 30\% exporters. We then run a regression of these countries' \rank{GDPpc} vs. \rank{Fitness}, and measure the Spearman correlation coefficient $R^2$. We finally average the coefficients per box. This figure shows the pattern formed by the indicator on the \theplane. It is clearly visible that products in the top left corners of the plane have exporters with high GDPpc, but both high and low Fitness. Losing low-Fitness exporters causes them to increase their Complexity, and this generates positive velocities along the horizontal axis. Products in the bottom right corner are much more homogeneous, having only high-Complexity exporters, and undergo less change in Complexity. This figure refers to the Feenstra dataset.}
    \label{fig:rexporters}
\end{figure}

\textbf{Beyond the Herfindahl approximation.}
Finally, a word on the discrepancy between Eq. \ref{eq:hfield-grad-def}'s predictions and the observed positive velocities along the Complexity axis in the top left part of the plane. This situation can be well understood in terms of the following reasoning. Products in this area have high GDPpc exporters, but both high and low Fitness exporters. When they move towards the diagonal, their exporters' GDPpc and Fitness tend to become more correlated to each other (through changes in the set of exporters). For some of them, this means losing low-Fitness exporters, and this causes an increase in Complexity, generating positive Complexity velocities in this area. logPRODY adjusts accordingly to the new Complexity value, but our model does not predict positive velocities because a high $H$ value only signals that the market shape is away from \asymptoticmarket, while a change in Complexity implies a change in the \asymptoticmarket\ value of a product. On the contrary, products in the lower right part of the plane tend to have high-Fitness exporters only, since the Complexity of the products found in this area is bound by their lowest Fitness exporter. It is extremely improbable for low-Fitness countries to start exporting high-Complexity products, and this cancels negative velocities on the Complexity axis in the area. To confirm this, we look at the exporter set of products. For each product $p$, we take its top 30\% exporters, as measured with \trca. We then run a regression of these countries' rank(GDPpc) vs. rank(Fitness), and measure the Spearman correlation coefficient $R^2$. A plot of the pattern formed by this indicator on the \theplane\ is shown in \figurename\ref{fig:rexporters}. It is clearly visible that products in the top left corners of the plane have exporters with high GDPpc, but both high and low Fitness, since correlation in these areas is lower. These results explain another aspect of a products' \asymptoticmarket: for a low-Complexity product, around the \asymptoticmarket, the Fitness level of its exporters is correlated to their GDPpc level. In order to capture this aspect of the dynamics on the \theplane, one needs to go beyond the information supplied by just $H$ and use additional data. \newline

In conclusion, the position and movement of the products on the plane is related to the configuration of their respective export markets. In general, market tend to the highest competition possible given the Complexity level of the product, and this high level of competition is associated with a characteristic position on the \theplane\ plane and a characteristic shape of the market, which we called ''asymptotic``, both depending on the Complexity level. Markets seem to randomly move away from the asymptotic state, and this is associated with movement away from the \asymptoticzone on the \theplane\ . The velocity field shows that there is a strong tendency to move back again to the asymptotic configuration of the market.

\section*{Discussion}

To summarise, the first finding is that there is a value of logPRODY products tend to move towards, which depends on their Complexity value. The value of logPRODY a product tends towards is related to its Complexity in a predictable way; this suggests that the Complexity of a product determines the way its global export market is shaped. Further analysis seems to confirm that there is an \asymptotic\ configuration for a product's export market that we called \asymptoticmarket, which we were able to characterize. High-complexity products have the simplest \asymptoticmarket, as they see countries with the highest Fitness develop the highest comparative advantage in making them. A more interesting finding is that,  at \asymptoticmarket, comparative advantage for lower Complexity products is not peaked on low Fitness countries, but seems instead to be, in general, more evenly distributed among all countries. On the plane there is an area that we called the \asymptoticzone, which is also characterized by the highest competition values, and the average motion of products tends towards it. Products on the \asymptoticzone\ have on average a particular export market configuration, that we called \asymptoticmarket, whose shape depends on their Complexity. Once products are in the \asymptoticzone, they can move away from it in a random fashion, because of a random noise effect. We attribute this effect to contingencies that change the shape of the market into something different from its \asymptoticmarket\ and, on average, decrease the competition. So in order to understand the motion on the \theplane, both this smooth average motion towards the \asymptoticzone\ and the random motion away from it need to be taken into consideration. As a result of the two, the full picture is self-consistently stationary, with products regularly going towards their \asymptoticmarket, and randomly exiting it.\newline

The comparison of monetary and non-monetary metrics proposed in the Economic Complexity framework, and the use of complexity tools such as dynamical system methods, allows us to observe regularities in the export markets' behavior that are hard to notice with more conventional mathematical tools. By looking at the trajectories of products on the \theplane, we are able to infer that there seems to be an \asymptotic\ value of logPRODY which depends on the level of Complexity, and whenever a product is away from its \asymptotic\ logPRODY it tends to move towards it. Study of the underlying markets showed that vertical movement in the \theplane\ is associated with moving from low-competition areas of the plane towards zones with higher average competition, and this happens with such regularity that we can use Eq.\ref{eq:hfield-grad-def} to describe the average motion. Therefore the change in logPRODY is caused by a shift in the set of countries that export a product at a given time: logPRODY can be seen as a synthetic index describing the export market for a product, whereas it cannot convey all the information given by a market representation such as those of \figurename \ref{fig:marketvectors}. The fact that the \asymptotic\ logPRODY for a product is determined for the most part by the corresponding value of Complexity suggests therefore that the shape of the export market is determined by the complexity of a product. Complexity, in turn, expresses how difficult it is to make a product, and hints to some bounds on which countries can make certain products.\\
Study of the coupled time evolution of the two indices is telling us that the logPRODY of a product is usually aligned with its Complexity level, meaning that the players in the product's export market are distributed in a certain typical way, that we called \asymptoticmarket. Whenever this does not happen, logPRODY tends to change to remove this misalignment. This change reflects a change in the market for the product, that tends to move towards an \asymptoticmarket situation. The Complexity level of a product determines how the \asymptoticmarket\ market is shaped, and in turn the way the market is shaped informs us on the Complexity level of a product. Higher Complexity products tend to have markets in which only high-Fitness countries develop a significant comparative advantage. On the other hand, lower-Complexity products do not show a peak on the lowest-Fitness countries: a ``flat'' distribution of comparative advantage is observed among all countries at \asymptoticmarket\ (even though some of these products exhibit a distribution with a slight peak, it is nowhere near the sharpness seen for complex products). The observed phenomena could be said to be a process of return to equilibrium on a short time scale, as there certainly are well behaved regularities in the aggregated behavior of products, consistent across all the data examined. On longer time scales, there is evidence that low-Complexity products tend to be replaced by high-Complexity products as countries develop\cite{Klimek2012}.\newline

We speculate that the observed \asymptoticmarket\ market shapes is a consequence of the triangular shape of the export matrix\cite{Tacchella2012}, which in turn follows from the fundamental empirical finding\cite{Hausmann2009} that development of a country grows together with diversification, from low-Complexity export to more sophisticated products. We remind that the export matrix is the adjacency matrix of the export network, connecting a country $c$ to all the products $\{p\}$ that it exports. High-Complexity products are exported only by a few extremely diversified countries, that also tend to have very high Fitness, and this could be the cause of the narrow peak of comparative advantages observed in \figurename \ref{fig:marketvectors}. Products with lower Complexity, instead, are exported by almost all countries, and not just by low-Fitness ones; this suggests an explanation for why, at \asymptoticmarket, we generally see flat or gently peaked market shapes for this kind of products. Because of the shape of the adjacency matrix, and the definition of Complexity, there is an approximate maximum level of competition for a product. This maximum decreases with the Complexity of the product, because there are few countries that can export high-Complexity products. Our finding is the observation that, when at this maximum of competition, the market is shaped in a characteristic way, that also depends on the Complexity, and that we called \asymptoticmarket. \newline

Furthermore the dynamics out of the asymptotic regime raises a natural question concerning the creation of new opportunities or movements driven by the attempt of reducing the competition or the creation of commercial niche: roughly speaking, whether schemes like red ocean/blue ocean \cite{kim2005blue} can be somehow mapped into this kind of analysis and, in general, whether they would applicable to bilateral trade networks. However, moving forward in this line of reasoning would require reliable and granular product value added data and not simply raw measures of product value. As previously mentioned, there is a general scarcity of value added data. We basically find value added data only for US internal production at a level of aggregation comparable to HS2007 4 digits. Using only this data would require the clearly poor assumption that all countries share the same structure of product value added. On the other hand, we have other sources of data which cover approximately 50-60 countries and approximately one decade (mainly OECD countries plus few large developed non OECD countries) but unfortunately they are defined at a much more aggregated level (comparable with 1 or 2 digit level of HS2007). Including value added data represents one of the challenging next steps of our analyses and the latter data-sets are clearly the most promising candidates but they require additional modeling in order to break down aggregate value added at a more granular level. 

Finally, a finding in itself is that results obtained with the Herfindahl field $H$ suggest that models inspired by statistical physics and complexity science can be used effectively to explore and understand the behavior of economic phenomena. Just regressing the value of $H$ versus the value of logPRODY cannot take into account the dependency on Complexity and changes in time, resulting in a correlation coefficient near to zero. Measuring the average value per box of an observable on the \theplane, and representing it as a scalar field, is a useful extension of the Economic Complexity toolbox, that could be applied in the future in to shed light on other processes being examined, where more traditional approaches might fail. We also believe that the great consistency of results across different datasets speaks to the coherence and well-groundedness of our methods, and the Economic Complexity framework as a whole.

\section*{Methods}

Here we add some information on the data and methodologies needed for the reproduction of our results. This section is organized into sub-sections. In this order, we will illustrate the datasets used and their structure, clarify the meaning of the non-monetary Fitness and Complexity metrics, explain the role of the ranking function in our analysis, and finally make some remarks on the regression techniques we used.
\subsection*{Datasets}
In this work we use two different datasets, that contain all the information of the $\exm_{cp}$ matrix (from which all the metrics considered in this paper can be calculated, except for GDP). We will call the first dataset BACI, and it is described in\cite{Gaulier2010}. It consists in data obtained by the UN-COMTRADE, and elaborated by CEPII, from which it has been purchased. For this reason, it cannot be made publicly available, although a free version without data cleaning is available on the BACI section of the organization's website\cite{datasetBACI}. We use the data for 148 countries, and spanning the 20 years from 1995 to 2014. In this dataset, products are classified according to the Harmonized System 2007\cite{datasetWorldCustomsOrg} which denotes them with a set of 6-digit codes organized in a hierarchical fashion. The code is divided into three 2-digit parts, each specifying one level of the hierarchy: so the first part indicates the broadest categories, such as "live animals and animal products" (01xxxx) or "plastics and articles thereof" (39xxxx). The second two digits specify further distinctions in each category, such as "live swine" (0103xx), and "live bovines" (0102xx). The last two digit are even more specific. For the analysis mentioned in the paper, we look at data for products aggregated at 4-digit level (1131 products). For the analysis on prices, for which we do not mention any results, we use a similar dataset, again released by BACI, containing all the transactions between countries in the 2008-2014, at 6-digit level. The second dataset for which we show results  in the paper is available on the NBER-UN world trade data website\cite{datasetNBER}, extensively documented by Feenstra et al. in \cite{Feenstra2005}, therefore we call it the Feenstra dataset in this work. After data cleaning procedures, we retain the data for 158 countries and 538 products, classified according to the SITC rev.2\cite{datasetSITC2}. Data cleaning procedures have been performed with the same methodologies on both datasets for which we present results. The procedures consist in the elimination of extremely small countries and countries with fragmented data; the aggregation of some product categories that are closely related, and a regularization of the $M_{cp}$ matrices. The dataset spans 36 years, from 1963 to 2000. GDPpc data has been downloaded from the World Bank Open Data website \cite{datasetWorldBank}. All results shown in this paper are consistent across all the datasets examined.

\subsection*{Non-monetary metrics}
As already discussed, Fitness and Complexity measures stem from an attempt to improve current theories about economic growth. It has been shown that developed countries show a high diversity of export, while poorly diversified economies tend to be competitive in just those products that are exported by almost all other countries\cite{Hausmann2009}. A model to explain this phenomenon is given in \cite{Hausmann2007}: the bipartite products-countries network is actually the only measurable part of a tripartite products-capabilities-countries network. The capabilities are unobservable, and are linked to both the countries that possess them and the products that need them in order to be exported. A link is added between country $c$ and product $p$ if $c$ is linked to all capabilities ${k}$ needed to export $p$. This will result in a nested products-countries network, as some capabilities are common to almost all countries, and others are rarer, and associated to developed countries only. Since diversification appears to be crucial to development of an economy, Fitness and Complexity are designed to efficiently extract nestedness information from the bipartite network of exporting countries and exported products. This is obtained via an algorithm iterating a highly non-linear map onto the adjacency matrix of the export network. The map is designed according to the following two observations. First, a product being made by a very diversified country is uninformative, while if it is made by at least one underdeveloped country we have grounds to think that it is a low-Complexity one. Vice-versa, if a product is made only by high-Fitness countries, it is reasonable to expect that it has high Complexity. The result is one of the simplest algorithms capable to be coherent and compatible with the kind of information being manipulated, taking advantage of the highly nested nature of the export network. An important feature of the algorithm, which is important to our analysis, is that the Complexity of a product has an upper bound, namely the lowest Fitness value found among its exporters. We define $M$ as the adjacency matrix of such a network: $M_{cp}$ is equal to 1 if country $c$ is an exporter of product $p$ (defined as $\rca_{cp}\geq1$), and 0 if it is not ($\rca_{cp}<1$). Fitness and Complexity values are defined as the fixed point of this non-linear coupled map:

\begin{eqnarray}
F_c^{(0)} = 1 \quad \forall c, &\cpx_p^{(0)} = 1 \quad \forall p. \\
\tilde{F}_c^{(n)} = \sum_p M_{cp} \cpx_p^{(n-1)},\qquad  &\tilde{\cpx}_p^{(n)} = \frac{1}{\sum_c M_{cp} \frac{1}{F_c^{(n-1)}}}\\
F_c^{(n)} = \frac{\tilde{F}_c^{(n)}}{\langle \tilde{F}_c^{(n)} \rangle_c},  &\cpx_p^{(n)} = \frac{\tilde{\cpx}_p^{(n)}}{\langle \tilde{\cpx}_p^{(n)} \rangle_p}.
\label{eq:fitcompmap}
\end{eqnarray}

This defines an iterative process coupling the $F_c$ to the total complexities $\cpx_p$ of products exported by $c$. The equation for $\cpx$ is non-linear, and bounds the complexity of a product to be smaller than the lowest Fitness value found among its exporters. The iteration is run until the values of $F$ and $\cpx$ reach a fixed point. The fixed point can be defined numerically in various ways. Here we used the definition proposed by Pugliese et al.\cite{Pugliese2016}, based on the stability of the ranking, which has to stop changing for a high number of successive iterations. This algorithm is capable of capturing the characteristic nestedness of the bipartite export networks, which is relevant to examining the dynamics of industrialization found in developing countries, but also to analyze in detail the production structure of developed countries\cite{Zaccaria2016,Pugliese2015,Cristelli2014m}. Further details can be found in \cite{Tacchella2012,Cristelli2013a}, the convergence problem is studied in \cite{Pugliese2016,Wu2016}, and a study of the stability of these metrics with respect to noise in the data has been performed in \cite{Battiston2014,Mariani2015}.

\subsection*{Monetary metrics}
Both aforementioned Sophistication and PRODY for a product $p$ are defined as a suitably weighted average of the GDPpc's of $p$'s exporters. In Sophistication, the weights are the export market shares of countries for product $p$, while PRODY uses the \trca\ values, defined in the following. We found that results are essentially similar whether one uses PRODY or Sophistication, as some literature seems to already suggest\cite{Minondo}. We identify logPRODY as the monetary metric of choice, which is a modification of the PRODY index proposed by Hausmann\cite{Hausmann2007}, who employed it to investigate the relationship between exports and growth of a country. logPRODY is defined, for a product $p$, as follows:
\begin{equation}
\lpd_p \equiv \sum_c \frac{\rca_{cp} \log_{10}(\gdp_c)}{\sum_c \rca_{cp}} = \sum_c \nrca_{cp} \log_{10}(\gdp_c)
\label{eq:logprodydef}
\end{equation}
where $\rca_{cp}$  is the so-called Revealed Comparative Advantage (\trca), or Balassa index\cite{Balassa1965}, and we defined the weighs $\nrca_{cp} = \rca_{cp} / \sum_c \rca_{cp}$. If we define the value in dollars of product $p$ exported by country $c$ as $\exm_{cp}$, then the $\rca_{cp}$ is defined as:
\begin{equation}
 \rca_{cp} = \frac{\frac{\exm_{cp}} {\sum_j \exm_{cj}}} {\frac{\sum_i \exm_{ip}} {\sum_{kl} \exm_{kl}}}.
\end{equation}
 The original PRODY is defined the same way, except that $\log_{10}(\gdp_c)$ is replaced by $\gdp_c$ in the sum. The change to logarithms has been chosen because GDPpc's of countries span about four orders of magnitude, and the geometric mean is better suited to represent such a numeric distribution of values. Normalization of the weight $\nrca_{cp}$ takes away the dependency of \trca\ on the denominator in its definition. Therefore $\nrca_{cp}$ is proportional to $\exm_{cp}$ divided by the total export of country $\sum_j \exm_{cj}$, i.e. the share of $p$ in all of $c$'s exports. This means that the biggest the share of a certain product $p$ in the total export of a country $c$, the more $c$ is weighted in the average that defines $\lpd_p$. As with all economic measures, logPRODY has its own pitfalls. It is less reliable when used to characterize products whose production is location-specific, such as raw materials, some kinds of vegetables. An interesting example is live swine, which is very common in rich western countries, and almost absent throughout the islamic world. As a result, it is one of the products with the highest logPRODY. We ran all our calculations both with and without these peculiar products, and didn't notice any significant change in the signal, therefore we believe this shortcoming of the metric does not impact negatively on our findings. 
In this respect, we stress that the non linearity of the Complexity measure solves this kind of issues, because it is enough to have only one low fitness country that export a raw material to obtain a low complexity score for that product. These intrisic differences make the comparison of the two quantities particularly meaningful.

\subsection*{Ranking}
The coordinates of choice for exploring the dynamics of products are not directly the value of Complexity and logPRODY, but their yearly tied ranking (normalized to 1, with zero being the lowest ranking). This is because of a known phenomenon observed during the convergence of the map in Eq.\ref{eq:fitcompmap} defining Fitness and Complexity, namely that some values tend asymptotically to zero\cite{Pugliese2016}. At the numerically estimated fixed point of the map, Complexity values span about 300 orders of magnitude, from $10^1$ to $10^{-300}$, this number being limited only by computer precision. Some of the products reach complexities so low that they have to be approximated to 0. Additionally, the yearly change depends on the order of magnitude of the Complexity, i.e. a complexity of about $10^{-100}$ has typically changes of about $10^{-100}$, while a complexity of order 1 moves of about order 1. This makes it hard to compare velocities across the plane, even when it is represented with logarithmic proportions. Ranking has the nice property of assigning equal spacing between one value and the next, solving all of these problems. It also introduces potential distortions in the signal, though, as a product's rank is directly dependent on all other products' Complexity values. In order to evaluate this potential distortion, we study a model in which we rank a number of values that evolve in time with a random walk. We find that the distortions introduced by the ranking are progressively higher as one moves towards the edges of the ranking, and depend on the process underlying the random walk. In the case of our dataset's particular kind of motion, we find that the distortions introduced by the ranking are negligible in size.

\subsection*{Regression technique}
The main regression technique used throughout this work consists in gathering information, in the form of a field, from a series of discrete trajectories on a plane. To measure velocity, for example, one can consider all the positions associated to the trajectories as points on the plane. To each point a velocity is associated by considering how the point moved along the trajectory it belongs to. One can then divide the plane into boxes with a square grid, and average all the velocities of points in the same box, obtaining a discrete vector field. The same can be done for other observables, i.e. for the Herfindahl index one can associate to each point (representing the state of a product at a certain time) the corresponding value of $H$, and then make an average per box obtaining a scalar vector field. It is fundamentally a non-parametric regression technique, based on Lorenz's method of analogues\cite{Lorenz1969}. We checked that results presented in this paper hold using two more regression techniques, namely a Nadaraya-Watson kernel regression\cite{Nadaraya1964} and the \emph{bdynsys} R package by Ranganathan et al.\cite{Ranganathan2014}, which is essentially a utility for finding the best function to fit on the data among all the possible polynomials up to a certain degree. Both produce fields that are very similar to those obtained with the average box velocity estimation, and comparable results in all cases.

\section*{Acknowledgements}
We thank Fabio Saracco for performing some of the data cleaning procedures on the BACI dataset. We also thank Andrea Tacchella, for relevant discussions and suggestions on measurement of the distortions introduced by the ranking function. Finally, thanks to the CNR Progetto di Interesse CRISIS LAB (\texttt{http://www.crisislab.it}) and EU Project nr. 611272 GROWTHCOM (\texttt{http://www.growthcom.eu}).

\section*{Author contributions statement}
All authors contributed equally.

\section*{Additional information}
\textbf{Competing financial interests:} The authors declare that no competing interests exist. \textbf{Funding:} CNR Progetto di Interesse CRISIS LAB (\texttt{http://www.crisislab.it}) and EU Project nr. 611272 GROWTHCOM (\texttt{http://www.growthcom.eu}) covered the salaries of O.A., M.C. and A.Z., and the purchase of the datasets used. The funders had no role in study design, data collection and analysis, decision to publish, or preparation of the manuscript.
\textbf{Data Availability:} The raw data of the BACI dataset about export flows cannot be made publicly available because this dataset has been purchased from CEPII (\verb!http://www.cepii.fr/CEPII/en/bdd_modele/presentation.asp?id=1!). However, the use of this dataset is not exclusive and anyone can purchase it. A free version for academic and research institutions is available on the United Nations COMTRADE website but differently from CEPII dataset, these data are missing from data sanitation. The details of the data sanitation are public and reported in reference \cite{Gaulier2010} of the present paper. To ask questions about accessing the data, please contact baci@cepii.fr. The Feenstra dataset is available at reference \cite{Feenstra2005} of the present paper.

%
%
%

\end{document}